\documentclass[a4paper,11pt]{article}
\usepackage{jheppub} % for details on the use of the package, please see the JINST-author-manual
\makeatletter
\renewcommand\@fpheader{} % removes "Prepared for submission to JHEP"
\makeatother
\usepackage{float}
\usepackage{booktabs}
\usepackage{setspace}
\usepackage{subfigure} % Add this to your preamble if not already included.
\newcommand{\mbH}{\mathbb{H}}

\newcommand{\mA}{\mathcal{A}}

\newcommand{\mD}{\mathcal{D}}
\newcommand{\mO}{\mathcal{O}}

\newcommand{\mZ}{\mathcal{Z}}

\newcommand{\be}{\begin{equation}}
\newcommand{\ee}{\end{equation}}
\newcommand{\bea}{\begin{eqnarray}}
\newcommand{\eea}{\end{eqnarray}}
\newcommand{\ba}{\begin{align}}
\newcommand{\ea}{\end{align}}
\newcommand{\bse}{\begin{subequations}}
\newcommand{\ese}{\end{subequations}}
\newcommand{\comment}[1]{}

\arxivnumber{2403.17149} % if you have one

\title{\boldmath Symplectic Quantization: numerical results for the Feynman propagator on a 1+1 lattice and the theoretical relation with Quantum Field Theory}

\author[a,b]{Martina Giachello}
\author[c,d]{Francesco Scardino}
\author[a,b]{Giacomo Gradenigo}

\affiliation[a]{Gran Sasso Science Institute, Viale F. Crispi 7, 67100 L'Aquila, Italy}
\affiliation[b]{INFN-Laboratori Nazionali del Gran Sasso, Via G. Acitelli 22, 67100 Assergi (AQ), Italy}
\affiliation[c]{Physics Department, INFN Roma1, Piazzale A. Moro 2, Roma, I-00185, Italy}
\affiliation[d]{Physics Department, Sapienza University, Piazzale A. Moro 2, Roma, I-00185, Italy}

\emailAdd{martina.giachello@gssi.it}

\abstract{We present here the first lattice simulation of symplectic quantization, a new functional approach to quantum field theory which allows to define an algorithm to numerically sample the quantum fluctuations of fields directly in Minkowski space-time, at variance with all other present approaches. Symplectic quantization is characterized by a Hamiltonian deterministic dynamics evolving with respect to an additional time parameter $\tau$ analogous to the fictious time of Parisi-Wu stochastic quantization. The difference between stochastic quantization and the present approach is that the former is well defined only for Euclidean field theories, while the latter allows to sample the causal structure of space-time. In this work we present the numerical study of a real scalar field theory on a 1+1 space-time lattice with a $\lambda \phi^4$ interaction. We find that for $\lambda \ll 1$ the two-point correlation function obtained numerically reproduces qualitatively well the shape of the free Feynman propagator. Within symplectic quantization the expectation values over quantum fluctuations are computed as dynamical averages along the dynamics in $\tau$, in force of a natural ergodic hypothesis connecting Hamiltonian dynamics with a generalized microcanonical ensemble. Analytically, we prove that this {\it microcanonical} ensemble, in the continuum limit, is equivalent to a {\it canonical-like} one where the probability density of field configurations is $P[\phi] \propto \exp(zS[\phi]/\hbar)$. The results from our simulations correspond to the value $z=1$ of the parameter in the canonical weight, which in this case is a well-defined probability density for field configurations in causal space-time, provided that a lower bounded interaction potential is considered.  The form proposed for $P[\phi]$ suggests that our theory can be connected to ordinary quantum field theory by analytic continuation in the complex-$z$ plane.}

\begin{document}
\maketitle
\flushbottom

\tableofcontents
\section{Introduction}
\label{intro}
Since its invention by Kenneth Wilson~\cite{W74}, lattice field theory
had an enormous development~\cite{C83,MM94} as a method to handle
non-perturbative problems in quantum field theory, in particular
concerning the theory of strong interactions with respecto to problems such as
the estimate of hadronic masses~\cite{FH12} or heavy ions
collisions~\cite{R18}. Nevertheless, despite the great achievements,
any numerical approach to quantum field theory on the lattice has retained so far a major limitation: all importance-sampling protocols are well defined
only for Euclidean field theory. Since the latter is obtained by Wick-rotating real into imaginary time, the causal structure of correlation functions in quantum field theory cannot be usually sampled numerically. The convenience/necessity to analytically continue real to immaginary time is to transform the Feynman path integral, characterized by the
oscillating factor $\exp(i S[\phi]/\hbar)$ ~---~ $S[\phi]$ is the relativistic action and $\phi$ a generic quantum field ~---~into a normalizable probability density $\exp(- S_E[\phi]/\hbar)$,
with $\hbar$ playing the same role of temperature in the Boltzmann
weight of statistical mechanics and where $S_E[\phi]$ is the positive-definite Euclidean action. The mapping to imaginary time has been so far the unavoidable condition to set up any importance sampling numerical protocol to study the quantum fluctuations of
fields. The goal of the present work is to present a conceptual
framework and a method which goes beyond this limitation, allowing for a
straightforward procedure to numerically sample quantum fluctuations of
fields in Minkowskian spacetime. The use of imaginary
time and Euclidean field theory forbids the representation on the
lattice of any process or phenomenon intrinsically related to the
causal structure of space-time, in particular all processes on the
light cone. By definition the probability density $\exp(-
S_E[\phi]/\hbar)$ works as an effective ``equilibrium'' measure for quantum
fluctuations. Any importance-sampling protocol built from the Euclidean weight projects, for large lattice sizes, on the ground states of the corresponding Minkowskian theory. In fact, Monte Carlo simulations built from Euclidean field theory allow us to reproduce with extreme precision the physics of 
{\it stable/equilibrium} bound states of the strong interactions~\cite{FH12}, whereas it has been so far much more problematic to reproduce metastable resonances with short lifetimes, like for instance tetraquark or pentaquark states~\cite{MPR15,EPP17}, or the dynamics of scattering processes with a strong relativistic character, namely processes with a different number of degrees of freedom in the asymptotic initial and final states. It is for this reason that we believe it is of crucial interest the possibility to test numerically any new proposal for a quantum field theory formulation which allows first to define and then to study the dynamics of quantum fields fluctuations directly in Minkowski
space-time.\\
An interesting idea in this direction, namely the proposal of a
functional approach to field theory which is well defined from the
probabilistic point of view already in Lorentzian space-time, has been
recently put forward by one of us and goes under the name of {\it
  ``symplectic quantization''}~\cite{GL21,G21,GLS24}. The first ingredient of this approach is to assume, for a given quantum field $\phi(x)$, with $x = (ct,{\bf x})$
a point in four-dimensional space-time, a dependence on an
additional time parameter $\tau$:
\begin{align}
\phi(x) ~\rightarrow~\phi(x,\tau),
\end{align}
which controls the continuous sequence of quantum fluctuations in each
point of space-time. Theories with such an additional time parameter
are not a novelty, the whole Parisi-Wu stochastic quantization
approach being based on this idea~\cite{PW81,DH87}.

Within the stochastic quantization approach,
Euclidean multipoint correlation functions are obtained as
a time average over the fictitous time parameter, 
\begin{align}
\lim_{\tau\rightarrow\infty} \frac{1}{\tau} \int_{0}^\tau \phi(x_1,s)\ldots\phi(x_n,s) = \langle \phi(x_1)\ldots\phi(x_n)\rangle_E,
\end{align}
where on the right hand term of the above equation $\langle
~\rangle_E$ denotes expectation with respect to the Euclidean weight
and on the left hand term each $\phi(x_i,s)$ represents the solution
at point $x_i$ of an auxiliary Langevin dynamics of the kind
\begin{align}
\frac{d\phi}{d\tau} = - \frac{\delta S_E[\phi]}{\delta \phi(x,\tau)} +
\eta(x,\tau),
\label{eq:Langevin}
\end{align}
where $\eta(x,\tau)$ represents is a zero mean white noise. It is a standard result
in the theory of stochastic processes \cite{R96} that the
stochastic dynamics of Eq.~\eqref{eq:Langevin} allows to sample
asymptotically the equilbrium distribution $\exp\left(
-S_{E}[\phi]/\hbar \right)$.  Shortly after the seminal idea of
stochastic quantization, people realized that, on the basis of
statistical ensemble equivalence, the Euclidean probability density of
quantum fields can be sampled also by following the solutions of the
Hamilton equations generated by a generalized Hamiltonian functional
$\mathbb{H}_E[\pi(x),\phi(x)]$ of the kind
\begin{align}
\mathbb{H}_E[\pi(x),\phi(x)] = \mathbb{K}_E[\pi(x)] + S_E[\phi(x)],
\end{align}
where $\pi(x,\tau)$ is a generalized momentum conjugated to the field
$\phi(x,\tau)$ with respect to the flowing of the fictious time
$\tau$. The idea of~\cite{CR82,DFF83}, which goes under the name of {\it
  ``Microcanonical approach to quantum field theory''}, is to achieve a sampling of
the Euclidean probability of the fields by studying the following deterministic
equations:
\begin{align}
  \dot{\phi}(x,\tau) &= \frac{\delta }{\delta\pi(x)}\mathbb{H}_E[\pi,\phi]\nonumber \\
  \dot{\pi}(x,\tau) &= -\frac{\delta }{\delta\phi(x)}\mathbb{H}_E[\pi,\phi].
\label{eq:Ham-Eu}
\end{align}
The replacement of the stochastic dynamics in Eq.~\eqref{eq:Langevin}
with the deterministic one in Eq.~\eqref{eq:Ham-Eu} was solely
motivated by its major computational efficiency in certain specific
situations, for instance in the case of non-local bosonic actions
obtained from the integration of fermionic variables~\cite{MM94},
where the deterministic equations are more suited to parallel updates
of the variables. As such, the role of the microcanonical ensemble
built on the the conservation of the generalized Hamiltonian
$\mathcal{H}_E[\pi(x),\phi(x)]$ was merely that of an alternative
technique to sample $\exp\left( -S_{E}[\phi]/\hbar \right)$, where no
physical interpretation was given neither to the additional time
$\tau$ nor to the conjugated momenta $\pi(x,\tau)$. On the contrary,
the key idea of symplectic quantization is to claim that the
microcanonical approach to quantum field theory is something more
fundamental and general, valid independently from its formal
equivalence to the Euclidean field theory: this, as we are going to
show, allows to sample quantum fluctuations directly in Minkowskian
space-time. The logic of the following exposition will be therefore
precisely the opposite of the one used to introduce the microcanonical
approach to Euclidean quantum field theory, justified solely on its
formal equivalence with the latter. First, without knowking which is
the corresponding canonical ensemble, we claim the existence of a
microcanonical ensemble built on a generalized Hamiltonian of the kind
\begin{align}
\mathbb{H}[\pi(x),\phi(x)] = \mathbb{K}[\pi(x)] - S[\phi(x)],
\end{align}
where again we have a generalized kinetic energy term
$\mathbb{K}[\pi(x)]$ and where this time the generalized potential is
$- S[\phi(x)]$, with $S[\phi(x)]$ the original Minkowskian action. We
will show that the two point correlation function obtained by
generating the quantum fluctuations of a weakly non-linear $\lambda
\phi^4$ theory from the dynamics generated by
$\mathbb{H}[\pi(x),\phi(x)]$ looks precisely like the standard Feynman
propagator. It will be only after having checked that the
deterministic dynamics generated by $\mathbb{H}[\pi(x),\phi(x)]$
allows to reproduce results in qualitative agreement to ordinary
quantum field theory that we will find out which is the canonical
weigth corresponding to our deterministic dynamics, obtaining it as
the main result of the Sec.~\ref{sec:pertequiv}. In particular, in
Sec.~\ref{sec:pertequiv} we will show that in the thermodynamic limit
the sampling of quantum fluctuation with our symplectic dynamics is
equivalent to the sampling according to a canonical weight of the kind
\begin{align}
    P_z[\phi] \propto \exp\left( z S[\phi]/\hbar \right),
    \label{eq:canonical-prop-1}
\end{align}
where the results of our simulations correspond to fixing $z=1$ in
Eq.~\eqref{eq:canonical-prop-1}. Clearly, a probability density like
the one in Eq.~\eqref{eq:canonical-prop-1} is not equivalent to the
complex amplitude which enters the Feynman path integral, but can be
related to it by analytic continuation in the complex $z$
plane. Investigations in this direction are in progress and will be
presented in a forthcoming paper explaining how the symplectic
quantization approach works for a quantum particle in a harmonic
potential, with particular emphasis on how the analytic continuation
to complex $z$ plane of the above density,
Eq.~\eqref{eq:canonical-prop-1}, must be handled~\cite{GSG25}.\\

In synthesis, main idea of the symplectic quantization
  approach is the possibility to sample quantum fluctuation directly
  in Lorentzian space-time by means of a generalized microcanonical
  ensemble, built by adding the conjugated momenta with respect to the
  intrinsic time $\tau$ {\it directly} to the relativistic action with
  Minkowski signature, with no need of considering any sort of
  analytic continuation from real to immaginary time and therefore
  preserving the causal structure of space-time. This intuition has
  been strongly inspired from the evidence that in statistical
  mechanics there are physically important situations where only the
  microcanonical ensemble is well defined~\cite{GILM21,GILM21b},
  namely physical phenomena which cannot be described within the
  canonical ensemble.\\

\section{Symplectic Quantization: from dynamics to ensemble averages}
\label{sec:SQ-foundations}
Let us summarize here the main steps for the derivation of the
symplectic quantization dynamics. First of all, inspired by the
stochastic quantization approach~\cite{PW81,DH87}, we assume that
quantum fields $\phi(x,\tau)$ depend on an additional time variable
$\tau$ which parametrizes the dynamics of quantum fluctuations in a
given point of Minkowski space-time. Since for a relativistic
quantum field theory the ambient space includes observer's time,
necessarily the intrinsic time $\tau$ must be a different variable, as
thoroughly discussed in~\cite{GL21,G21}. The symplectic quantization
approach to field theory assumes, consistently with the existence of
an intrinsic time $\tau$, the existence of conjugated momenta of the
kind
\begin{align}
  \pi(x,\tau) \propto \dot{\phi}(x,\tau),
\end{align}
which can be obtained as follows. First, we introduce a generalized
Lagrangian of the kind
\begin{align}
  \mathbb{L}[\phi,\dot{\phi}] = \int d^dx~\left[ \frac{1}{2c_s^2}\dot{\phi}^2(x) + S[\phi] \right],
\end{align}
where $c_s$, in natural units, is a dimensionless parameter, and $S[\phi]$ is
the standard action for a quantum field, e.g.,
\begin{align}
   S[\phi] &= \int d^dx \left( \frac{1}{2}\partial_\mu\phi(x)\partial^\mu\phi(x) - V[\phi(x)] \right) \nonumber \\
  & = \int d^dx \left[ \frac{1}{2}\left(\frac{\partial\phi}{\partial x^0}\right)^2 -\frac{1}{2}\sum_{i=1}^{d} \left(\frac{\partial\phi}{\partial x^i}\right)^2   - V[\phi(x)] \right] 
\end{align}
where the potential is, for instance
\begin{align}
V[\phi]= \frac{1}{2}m^2\phi^2+ \frac{1}{4}\lambda \phi^4.
\end{align}
By means of a Legendre transform one then passes to the Hamiltonian:
\begin{align}
   \mathbb{H}[\phi,\pi] &= \frac{1}{2} \int d^dx~c_s^2~\pi^2(x) - S[\phi]  \nonumber \\
  & = \int d^dx \left[ \frac{c_s^2}{2}\pi^2(x) - \frac{1}{2}\left(\frac{\partial\phi}{\partial x^0}\right)^2 + \frac{1}{2}\sum_{i=1}^{d} \left(\frac{\partial\phi}{\partial x^i}\right)^2 + V[\phi] \right]  \nonumber \\
  & = \int d^dx~\left[ \frac{c_s^2}{2}\pi^2(x) + \frac{1}{2}\phi~\partial_0^2\phi - \sum_{i=1}^{d} \phi~\partial_i^2\phi  + V[\phi]  \right]
\label{eq:sq-Hamiltonian}
\end{align}
For both the ease of notation and for conceptual simplicity we will assume hereafter $c_s=1$. Let us remark that there is no apriori reason forcing the new constant $c_s$ to be precisely equal to the speed of velocity. Yet, the fact that in natural units $c_s$ is dimensionless suggests that no new physical constants need to be introduced. From Eq.~(\ref{eq:sq-Hamiltonian}) we have that, within the symplectic
quantization approach, the dynamics of quantum fluctuations is the one governed by the following Hamilton equations:
\begin{align}
  \dot{\phi}(x) &= \frac{\delta \mathbb{H}[\phi,\pi]}{\delta\pi(x)} \nonumber \\
  \dot{\pi}(x) &= - \frac{\delta \mathbb{H}[\phi,\pi]}{\delta\phi(x)} = -\frac{\delta \mathbb{V}[\phi]}{\delta \phi(x)}, 
\label{eq:sq-Ham-eq}
\end{align}
from which one gets
\begin{align}
  \ddot{\phi}(x,\tau) = - \partial_0^2\phi(x,\tau) + \sum_{i=1}^{d} \partial_i^2\phi(x,\tau) - \frac{\delta V[\phi]}{\delta \phi(x,\tau)}.
  \label{eq:eq-motion}
\end{align}
At this stage one can legitimately wonder how a classical
deterministic theory can account for quantum fluctuations. Let us
notice that in the expression of the generalized Hamiltonian
$\mathbb{H}[\phi,\pi]$ we can recognize a {\it ``generalized potential
  energy''} $\mathbb{V}[\phi]$, corresponding to the original
relativistic action, and a {\it ``generalized kinetic energy''}
$\mathbb{K}[\pi]$, namely the quadratic part related to the new
conjugated momenta:
\begin{align}
  \mathbb{H}[\phi,\pi] &= \mathbb{K}[\pi] + \mathbb{V}[\phi] 
   \label{eq:H-separable1}
  \end{align}
  with:
  \begin{align}
  \mathbb{V}[\phi] & = - S[\phi] \nonumber \\
  \mathbb{K}[\pi] & = \frac{1}{2} \int d^dx~\pi^2(x).
  \label{eq:H-separable2}
\end{align}
The classical field $\phi_{cl}$ solution corresponds to a minimum of the new generalized potential
$\mathbb{V}[\phi]$:
\begin{align}
    -\frac{\delta \mathbb{V}[\phi]}{\delta \phi(x)} \bigg|_{\phi_{cl}}= 0.
\end{align}
On the contrary, quantum fluctuations are naturally sampled by the generalized Hamiltonian dynamics, along which the functional derivative of $\mathbb{V}[\phi]$ is not zero but equal to the rate of change of the conjugated momenta, see Eq.~\eqref{eq:sq-Ham-eq}. The generalized energy $\mathbb{H}[\phi,\pi]$ is constant along the dynamics: quantum fluctuations are encoded in the fluctuations of the generalized potential energy, $S[\phi]$.\\\\
Having defined the above deterministic dynamics for the quantum fluctuations of the field $\phi(x,\tau)$, Eq.~\eqref{eq:eq-motion}, one can then legitimately wonder how this functional
formalism connects to the standard one, for instance to the standard
Feynman path-integral formulation of quantum field theory. In first instance, 
in order to find out the connection between the dynamic approach of symplectic quantization and any functional formulation of field theory where the additional time $\tau$ is absent, one needs to find out which probability density $\rho[\phi(x)]$ corresponds to the dynamics. There are then two possibilities: $\rho[\phi(x)]$ might have either an integral or a punctual correspondence with quantum field theory. The integral correspondence is when the path-integral is recovered by means of an integral transformation, as it will be discussed below here, and the {\it punctual} correspondence is when, in a certain limit, one can directly show that $\rho[\phi(x)] \approx e^{i S[\phi]/\hbar}$. This last procedure is more subtle, because it requires an analytic continuation of the action and of its degrees of freedom in the complex plane, and will be discussed thoroughly in~\cite{GSG25,GSG25_2}. 
But let us now go back to the relation between symplectic dynamics and probability densities, for which we make an ergodic hypothesis for the Hamiltonian dynamics in Eq.~\eqref{eq:sq-Ham-eq}: if we assume that this dynamics samples at long time $\tau$ the constant generalized energy hypersurface with uniform probability~\cite{GL21,G21}, then we can associate to the dynamics of Eq.~\eqref{eq:sq-Ham-eq} the following measure:
\begin{align}
  \rho_{\text{micro}}[\phi(x)] = \frac{1}{\Omega[\mA]}~\delta\left(\mA-\mathbb{H}[\phi,\pi]\right),
  \label{eq:rho-micro}
\end{align}
where $\Omega[\mA]$ is a sort of {\it microcanonical} partition function
\begin{align}
  \Omega[\mA] = \int \mD\phi\mD\pi~\delta\left(\mA-\mathbb{H}[\phi,\pi]\right),
  \label{eq:part-micro}
\end{align}
with $\mD\phi = \prod_x d\phi(x)$ and $\mD\pi = \prod_x d\pi(x)$ the
standard notation for functional integration. From the above partition
function we can define the microcanonical adimensional entropy of
symplectic quantization:
\begin{align}
\Sigma_{\text{sym}}[\mA] = \ln \Omega[\mA]
\end{align}
The {\it ergodicity assumption} for the symplectic quantization
dynamics amounts to say that, considering $\mO[\phi(x)]$ a generic
observable of the quantum fields, symplectic quantization can be
related to a stationary probability measure free of the additional parameter $\tau$ by
claiming that for generic initial conditions the following equivalence
between averages holds:
\begin{align}
   \lim_{\Delta\tau \rightarrow \infty} \frac{1}{\Delta\tau}\int_{\tau_0}^{\Delta\tau} d\tau~\mO[\phi(x,\tau)] = 
   \int \mD\phi\mD\pi~\rho_{\text{micro}}[\phi(x)]~\mO[\phi(x)],
  \label{eq:mild-ergodicity}
\end{align}
%%%
where $\tau_0$ is a large enough time for the system to have reached
stationarity and ``lost memory'' of initial conditions. How to relate
then the microcanonical partition function in
Eq.~\eqref{eq:part-micro} to the path integral? It is quite intuitive
to understand that the two expression must be related by some sort of
statistical ensemble change. The crucial point of this change of
ensemble, as stressed already in~\cite{GL21}, is that the
microcanonical partition function $\Omega[\mA]$ is built on the
conservation of a non-positive quantity, the generalized Hamiltonian
$\mathbb{H}[\phi,\pi]$. The latter, from the point of view of physical
dimensions, is a relativistic action:  it therefore takes both arbitrarily large
positive and arbitrarily large negative values due to the negative sign in front of the
coordinate-time derivative term in the second line of
Eq.~\eqref{eq:sq-Hamiltonian}. The absence of positive definiteness
for the generalized Hamiltonian $\mathbb{H}[\phi,\pi]$, which is the
true relativistic signature of the theory, is what forbids a standard
change of ensemble with a Laplace transform, that is customary in
statistical mechanics when passing from microcanonical to canonical. The only integral transform which allows us to map {\it
  formally} the ensemble where $\mathbb{H}[\phi,\pi]$ is constrained
to the one where it is free to fluctuate is the Fourier transform.  It
is by Fourier transforming the microcanonical partition function
$\Omega[\mA]$ that one obtains straightforwardly the Feynman path
integral:
\begin{align}
  \mZ[u] & = \int_{-\infty}^\infty dA~e^{-iu\mA}~\Omega[\mA] = \int \mD\phi~\mD\pi~e^{- \frac{i}{2} u \int d^dx~\pi^2(x) + iu S[\phi]}  = \mathcal{N}(u) \int \mD\phi~e^{i u S[\phi]},
\label{eq:Feyn-path}
\end{align}
where $u$ is a variable conjugated to the action and in the second
line of Eq.~\eqref{eq:Feyn-path} we have integrated out momenta thanks
to the quadratic dependence on them, contributing the infinite
normalization constant $\mathcal{N}(u)$, which is typical of path
integrals. Finally, if we fix $u=\hbar^{-1}$ into the last line of
Eq.~\eqref{eq:Feyn-path} we have the Feynman path integral:
\begin{align}\label{eq:partition}
\mZ[\hbar] = \int_{-\infty}^\infty d\mA~e^{-iA/\hbar}~\Omega[\mA] \propto \int \mD\phi~e^{\frac{i}{\hbar} S[\phi]}.
\end{align}
The one above is to our knowledge the first derivation from first
principles of the Feynman path-integral formula in the context of a
more extended framework. We could say that this larger framework is
the statistical mechanics of action-preserving systems, opposed to the
statistical mechanics of energy-preserving systems, which is the
standard one.\\ 

At this stage one can therefore legitimately wonder which is
  the relation between the probability density of the extended
  framework just introduced, i.e., the $\rho_{\text{micro}}[\phi(x)]$
  of Eq.~\eqref{eq:rho-micro}, and standard quantum field theory
  probability amplitudes. The key to this, at the present stage of
  development of this new approach, is precisely the Fourier transform
  of Eq.~\eqref{eq:partition}: what we expect from the microcanonical
  symplectic quantization ensemble is the possibility to sample
  disconnected correlation functions at fixed generalized action $\mA$
  and then, by Fourier transforming with respect to $\mA$, obtain the
  original disconnected correlation functions of quantum field
  theory. We will see in Sec.~\ref{sec:Feynman} that remarkably good
  results for the two point correlation function can be obtained
  alredy in the fixed generalized action ensemble! We speak about
  disconnected correlation functions because, at the present stage of
  understanding, we are able to write down an explicit relation only
  between the generating functionals of disconnected correlations,
  respectively $\Omega[\mA]$ for symplectic quantization and
  $\mZ[\hbar]$ for quantum field theory. Clearly the new approach
  needs to be improved, for at least three main reasons. First, to
  recover disconnected correlation functions of quantum field theory
  by doing many deterministic simulations at different fixed values of
  action $\mA$ and then try to Fourier transform the result looks a
  quite impractical protocol. We are presently working on an attempt
  to improve the correspondence between symplectic quantization and
  quantum field theory, but is still in progress \cite{GSG25,GSG25_2}. Second, the physical
  information of a quantum field theory is contained in the connected
  correlators, not the disconnected ones, so to rebuild the whole
  theory from this path seems a very long way. Last, but not least, as
  we find in the result of simulations in Sec.~\ref{sec:stability} and
  we will derive analytically in Sec.~\ref{sec:pertequiv}, the
  microcanonical density of Eq.~\eqref{eq:rho-micro} is ill-defined
  for a free theory, which looks quite problematic when compared to
  the importance and the success of perturbative expansions in quantum
  field theory.\\

Let us now come back for a moment on the reason why,
  differently from what is customary in the theory of statistical
  ensembles, symplectic quantization is related to quantum field
  theory by Fourier rathern than Laplace transforming. This comes from
  the fact that the microcanonical statistical ensemble of sympletic
  quantization is built on the conservation of a non positive-defined
  quantity, landmark of the relativistic nature of the theory, that
  implies the necessity of Fourier transforming and determines the
fact that {\it locally} we can only access complex probability
amplitudes and not real probabilities. From the perspective of
symplectic quantization the replacement at the local level of
probabilities with probability amplitudes is therefore a direct
consequence of special relativity and a wise use of statistical
ensembles. To better understand this statement let us consider an
unrealistic situation where the symplectic action (generalized
Hamiltonian) $\mathbb{H}[\phi,\pi]$ was positive definite. In this
case one could change ensemble with Laplace rather than Fourier
transform,
\begin{align}
  \mZ[\mu]  = \int_{0}^\infty d\mA~e^{-\mu \mA}~\Omega[\mA]\propto \int \mD\phi~e^{\mu S[\phi]},
\end{align}
leading to a theory which is perfectly equivalent to standard
statistical mechanics in the canonical ensemble: locally there is a
probability density for the field configuration, $\rho(\phi) \propto
e^{\mu S[\phi]}$. We notice that the factor $\rho(\phi)$ is
intuitively well defined as a local probability density because for
typical configuration of the field, far from those corresponding to
ultrarelativistic particles, the relativistic action is usually negative
$S[\phi]<0$.\\
We have just shown how the standard path-integral formulation can be
recovered, on the basis of an ergodicity assumption, from the
symplectic quantization dynamics approach and which is the role played
by $\hbar$ within this, let us say, {\it change of ensemble}.  At the
same time it is not only legitimate but also necessary to wonder if
and how there is a {\it quantization constraint} involving $\hbar$
which can be imposed directly on the microcanonical ensemble of
symplectic quantization. The indication coming from the stochastic
quantization framework is that $\hbar$ must play a role analogous to
that of temperature. Therefore, as suggested in~\cite{GLS24}, we
believe that the most natural assumption for the role of $\hbar$ in
the symplectic quantization formalism is to be analogous to the
microcanonical temperature:
\begin{align}
  \frac{1}{\hbar} = \frac{d\Sigma_{\text{sym}}[\mA]}{d\mA}
  \label{eq:h-micro}
\end{align}
Although satisfactory conceptually and formally consistent, a
definition of $\hbar$ as in Eq.~\eqref{eq:h-micro} is very difficult
to implement in practice. For this reason we will resort in this paper
to another more trivial but effective way to impose the quantization
constraint in the symplectic quantization dynamics, the one analogous
to the way which is customarily used to assign the temperature in the
context a microcanonical molecular dynamics. Usually, if we have $M$
degrees of freedom and we wish the system to be on the fixed energy
hypersurface such that $T^{-1} = \partial S(E)/\partial E$, we simply
assign initial conditions such that the total energy is $E = M k_B T$:
here we follow the same strategy. In particular, counting as ``degrees
of freedom'' the number of components in reciprocal space of the
Fourier transform of the fields, i.e., $\pi(k)$ and $\phi(k)$, in
order to set at $\hbar$ the typical scale of generalized energy for
each mode we can choose initial conditions in the ensemble
characterized at stationarity by the following condition:
\begin{align}
  \langle \pi(x) \pi(y) \rangle = \frac{\hbar}{2}~\delta^{(d)}(x-y),
  \label{eq:corr-momenta-real}
\end{align}
where the angular brackets indicates intrinsic time average along the
symplectic quantization dynamics:
\begin{align}
\langle \pi(x) \pi(y) \rangle = \lim_{\Delta\tau \rightarrow \infty}
\frac{1}{\Delta\tau}\int_{\tau_0}^{\Delta\tau} d\tau~\pi(x,\tau)
\pi(y,\tau)
\label{eq:tau-dyn-av}
\end{align}
Eq.~\eqref{eq:corr-momenta-real} for the expectation value of momenta can be rewritten for a discretized $d$-dimensional
space-time lattice with lattice spacing $a$, as is the case for the numerical simulations discussed below, as follows:
\begin{align}
  \langle \pi(x_i) \pi(x_j) \rangle = \frac{\hbar}{2} ~\frac{\delta_{ij}}{a^d},
  \label{eq:corr-momenta-lattice}
\end{align}
where $\delta_{ij}$ is the Kronecker delta. By Fourier transforming
Eq.~\eqref{eq:corr-momenta-real} it is then straightforward to get
\begin{align}
  \langle \pi^*(k) \pi(k) \rangle = \frac{\hbar}{2},
  \label{eq:corr-momenta-Four}
\end{align}
so that in Fourier space the ``kinetic'' contribution coming from each
degree of freedom to the total action amounts to $\hbar/2$. The
relation in Eq.~\eqref{eq:corr-momenta-Four} can be also applied to
the discretized momenta usually considered for a numerical simulation on
the lattice:
\begin{align}
  \langle \pi^*(k_i) \pi(k_i) \rangle = \frac{\hbar}{2}\quad\quad
  \forall~i.
  \label{eq:corr-momenta-Four-lattice}
\end{align}
This will be the sort of quantization constraint which will be applied
to all our numerical simulations, choosing initial conditions which
are compatible with that. Since we have chosen to work with natural
units we will replace $\hbar=1$ everywhere in the above
formulas. Suitable initial conditions to expect something such as
Eq.~\eqref{eq:corr-momenta-Four-lattice} at stationarity is for
instance the following:
\begin{align}
 |\pi(k_i;\tau=0)|^2  = \hbar \quad\quad \forall~i,
  \label{eq:corr-momenta-Four-lattice-t0}
\end{align}
which will be used for all simulations presented in this work.\\

Before moving to the presentation of simulation details and
  the exposition of numerical results let us make a final remark on
  the assumption that dynamical averages along the intrinsic time
  Hamiltonian dynamics, like the one of Eq.~\eqref{eq:tau-dyn-av},
  yields indeed a reliable sampling of the microcanonical probability
  density in Eq.~\eqref{eq:rho-micro}. The general attitude in order
  to claim the equivalence between dynamical and ensemble sampling of
  observables expectation values is to assume a certain degree of
  ergodicity/chaoticity of the microscopic dynamics, which in general
  has to be considered case by case. To this respect, beside recalling
  that the case of a interacting scalar field theory with non-linear
  quartic potential studied here is generally considered the one of a
  system where the dynamics has good mixing properties, let us remark
  that in the present approach we do no really regard ergodicity as an
  issue. More precisely, we follow the perspective outlined in a
  recent review from one of us on the foundations of statistical
  mechanics \cite{BGVZ25} where it is explained how the use of statistical
  ensembles to draw predictions on the expectation values of generic
  observables requires ergodicity only in a weak sense, i.e., the
  mixing properties of the microscopic dynamics are not really the
  point. As for ``generic'' observables one has to think about all the
  observables which depend on large number of the system’s degrees of
  freedom. Just to make an example/analogy, we expect the behaviour of
  the two-point correlation function in real space for a weakly
  interacting field theory to ``thermalize well'' under the Hamiltonian
  dynamics in the same way that a good thermalization is achieved for
  individual particles in a classical harmonic chain, even if the
  Fourier modes of the harmonic chain does not thermalize at all. For
  further speculations on this point see, e.g., \cite{CVG22}.

%%%%%%%%%%%%%%%%%%%%%%%%%%%%%%%%%%%%%%%%%%%%%%%%%%%%%%%%%%%%%%%%%%%%%%%
%%%%%%%%%%%%%%%%%%%%%%%%%%%%%%%%%%%%%%%%%%%%%%%%%%%%%%%%%%%%%%%%%%%%%%%
%%%%%%%%%%%%%%%%%%%%%%%%%%%%%%%%%%%%%%%%%%%%%%%%%%%%%%%%%%%%%%%%%%%%%%%%%%%%%%%%%%%%%%%%%%%%%%%%%%%%%%%%%%%%%%%%%%%%%%%%%%%%%%%%%%%%%%%%%%%%%%
\section{Simulation details}
\label{sec:simulation}
The deterministic dynamics of symplectic quantization can be defined
for both Euclidean and Minkowski metric: to validate the new approach
we have tested both scenarios. In order to do that we have discretized
the Hamiltonian equations of motion, writing them in a general form
where the nature of the metric is specified by the variable $s=\lbrace
0,1 \rbrace$. All equations are written in natural units
$\hbar=c=1$.\\
In the present work we have considered a $1+1$ lattice with either
Euclidean or Minkowski metric, which we denote as $\Gamma$:
\begin{align}
  \Gamma&:\bigg\{x : x_\mu=an_\mu\,,\quad n_\mu=-\frac{M_\mu}{2},...,\frac{M_\mu}{2}\quad\mu=0,1 \bigg\}.
 \end{align}
Due to the finite size of the simulation grid momenta are also discretized:
\begin{align}
    p_\mu=\frac{2\pi}{a}\frac{k_\mu}{M_\mu}\quad\quad |p_\mu|<\frac{\pi}{a}, 
\end{align}
where $\mu=0,1$, $k_\mu \in [-M_\mu/2,M_\mu/2]$, $L=M_\mu a$ is the lattice side
and $a$ is the lattice spacing. In the discretized theory the total number of degrees of freedom
is identified with the number of points in the lattice:
\begin{align}
M = \prod_{\mu=0}^{d-1} M_\mu
\end{align}
The discretized Hamiltonian of
symplectic quantization reads then as
\begin{align}
  \mathbb{H}[\phi,\pi] = \frac{1}{2}\sum_{x \in \Gamma} \bigg[ \pi(x)^2  - (-1)^s \frac{1}{a^2} \phi(x)\Delta^{(0)}\phi(x) - \frac{1}{a^2} \phi(x)\Delta^{(1)}\phi(x)  
    + m^2\phi^2(x) + \frac{\lambda}{4} \phi^4(x) \bigg],
  \label{eq:discrete-Hamiltonian}
\end{align}
where the symbol $\Delta^{(\mu)}\phi(x)$ denotes the discrete
one-dimensional Laplacian along the $\mu$-th coordinate axis:
\begin{align}
\Delta^{(\mu)}\phi(x) = \phi(x + a^{\mu}) + \phi(x - a^{\mu})  - 2 \phi(x).
\end{align}
We have used a general expression in
Eq.~\eqref{eq:discrete-Hamiltonian}, which, depending on the value
chosen for the integer index $s=\lbrace 0,1 \rbrace$, describes a
theory with Euclidean, $s=0$, or Minkowskian, $s=1$, metric.  From the
expression of the Hamiltonian in Eq.~\eqref{eq:discrete-Hamiltonian}
we have that the force acting on the field on a two-dimensional
lattice is:
\begin{align}
  F[\phi(x)] &= - \frac{\delta\mathbb{H}[\phi,\pi]}{\delta \phi(x)} 
    = \frac{(-1)^s}{a^2} \Delta^{(0)}\phi(x) + \frac{1}{a^2} \Delta^{(1)}\phi(x) - m^2\phi(x) - \lambda \phi^3(x),  
  \label{eq:symp-force}
\end{align}
so that the equation of motion for the field itself is:
\begin{align}
  \frac{d\phi(x,\tau)}{d\tau^2} = F[\phi(x,\tau)].
  \label{eq:symp-motion}
\end{align}
%%
%% Let us now rewrite the equations of motion explicitly in terms of the
%% integer components $n_0$ and $n_1$ of $x = (a n_0,a n_1)$.  By
%% denoting $\phi(x,\tau) = \phi_{n_0,n_1}(\tau)$ we have:
%% %%
%% %%
%% \begin{align}
%%   & \frac{d{\phi}_{n_0,n_1}(\tau)}{d\tau^2} = \nonumber \\
%%   & (-1)^s\,\frac{\phi_{n_0+1,n_1}(\tau)+\phi_{n_0-1,n_1}(\tau)-2\phi_{n_0,n_1}(\tau)}{a^2}
%%   \nonumber \\
%%   & + \frac{\phi_{n_0,n_1+1}(\tau)+\phi_{n_0,n_1-1}(\tau)-2\phi_{n_0,n_1}(\tau)}{a^2} - \nonumber \\
%%   & - m \phi_{n_0,n_1}(\tau) - \lambda \phi_{n_0,n_1}^3(\tau) \nonumber \\
%%   \label{eq:discrete-scalar}
%% \end{align}
%% %%
%%
Equations~\eqref{eq:symp-force},\eqref{eq:symp-motion} define the
Hamiltonian dynamics which we have studied numerically using the
leap-frog algorithm, a symplectic algorithm described in
Appendix~\ref{appendix:algorithm}, which guarantees the conservation
of (generalized) energy at the order $\mathcal{O}(\tau^2)$.\\
An important point for the study of this paper is the definition of
boundary conditions. We used two different kinds of boundary conditions
for the simulations. For all results on Euclidean lattice and for the
study of dynamics stability with or without non-linear interaction on
Minkowski lattice, discussed respectively in Sec.~\ref{sec:euclidean}
and in Sec.~\ref{sec:stability}, we have used standard periodic
boundary conditions on the lattice. Differently, in
Sec.~\ref{sec:Feynman}, aimed at studying the free propagation of
physical signals across the lattice we used {\it fringe} boundary
conditions~\cite{NNH99}, introduced with the purpose of mimicking the
existence of an infinite lattice outside the simulation grid. Fringe
boundary conditions are realized considering a larger lattice, which
we denote as $\Gamma_{f}$, where the subscript {\it ``f''} is for
fringe, which is composed by the original lattice $\Gamma$ plus
several additional layer of points which we denote as
$\Gamma_{\text{ext}}$, in such a way that the {\it fringe} lattice is
$\Gamma_{f}=\Gamma+\Gamma_{\text{ext}}$. For the fringe lattice one
also considers periodic boundary conditions, but the generalized
Hamiltonian for points belonging to $\Gamma$ and to
$\Gamma_{\text{ext}}$ is different. Namely, the fringe lattice is
characterized by the Hamiltonian:
\begin{align}
  \mathbb{H}_{f}[\pi,\phi] = \mathbb{H}_{\text{ext}}[\pi,\phi] + \mathbb{H}[\pi,\phi],
  \label{eq:fringe-Hamiltonian}
\end{align}
where $\mathbb{H}[\pi,\phi]$ is the original discretized Hamiltonian
of the system, see Eq.~\ref{eq:discrete-Hamiltonian}, while
$\mathbb{H}_{\text{ext}}[\pi,\phi]$ reads as
\begin{align}
\mathbb{H}_{\text{ext}}[\pi,\phi]  = \frac{1}{2}\sum_{x \in \Gamma} \bigg[ \pi(x)^2  + 
  m^2\phi^2(x) + \frac{\lambda}{4} \phi^4(x) + \epsilon \left( \frac{1}{a^2} \phi(x)\Delta^{(0)}\phi(x)  - \frac{1}{a^2} \phi(x)\Delta^{(1)}\phi(x)\right)\bigg],
  \label{eq:discrete-Hamiltonian-fringe}
\end{align}
where the coefficient $\epsilon$ is very small, $\epsilon \ll 1$. This
choice of boundary conditions allows us to have a free propagation of
signals across the boundary layer of $\Gamma$, our true simulation
lattice, but the signal is then strongly damped when going across
$\Gamma_{\text{ext}}$, the ``external'' boundary layer before making
sort of interference at the periodic boundaries at the border of
$\Gamma_{\text{ext}}$. This choice of boundary conditions allows us
not only to deal with an overall system which is still Hamiltonian
(apart from small corrections scaling as $1/L$), but also to have
quite satisfactory results for the study of the Feynman propagator, as
shown in Sec.~\ref{sec:Feynman}.\\\\
We have done all simulations for a lattice with side $M_0=M_1=128$,
lattice spacing $a=1.0$ and using an integration time-step
$\delta\tau=0.001$. According to the discussion in the previous
section, we have fixed the energy scale by choosing initial conditions
such that each degree of freedom in Fourier space carries a {\it
  ``quantum''} of energy $\hbar=1$. We have therefore assigned an
initial total energy equal to $M_0\cdot M_1 = 16384$ for all
simulations. Since we have studied both linear and non-linear
interactions, in order to set precisely the initial value of the
energy, we started all simulations with:
\begin{align}
  & |\phi(k;0)|^2 = 0 \quad \quad \forall~k \nonumber \\ 
  & |\pi(k;0)|^2 = 1 \quad \quad \forall~k.
\end{align}

\section{Euclidean propagator}
\label{sec:euclidean}

Our first test of the symplectic quantization approach consists in the
study of its deterministic dynamics in the case of a two-dimensional
Euclidean lattice, showing that it provides the correct two-point
correlation function, also consistently with the results of stochastic
quantization. For the simulation on the Euclidean lattice we have used
simple periodic boundary conditions, since all correlation functions
decay exponentially with the distance and there should be no signals
propagating underdamped across the system.\\ Let us then recall here
how the expectation values over quantum fluctuations of fields are
computed within the symplectic quantization approach dynamics. If we
indicate with $\phi_{\mathbb{H}}(x,\tau)$ the solutions of the
Hamiltonian equations of motion written in Eq.~\eqref{eq:symp-motion},
we have that the expectation value of a generic $n$-point correlation
function can be computed as follows:
\begin{align}
  \langle & \phi(x_1), \ldots, \phi(x_n) \rangle  = \lim_{\Delta\tau\rightarrow\infty} \frac{1}{\Delta\tau} \int_{\tau_0}^{\tau_0+\Delta\tau} d\tau~\phi_{\mathbb{H}}(x_1,\tau)\ldots \phi_{\mathbb{H}}(x_n,\tau),
  \label{eq:def-correlations}
\end{align}
where $\tau_0$ is a large enough time, for which the system has
reached equilibrium and forgot any detail on the initial conditions of
the dynamics. For a free field theory the propagator on a
two-dimensional lattice take the simple form:
\begin{align}
\tilde{G}(p;a) = \left[ \frac{4}{a^2}\sin^2\left( \frac{a k_0}{2}\right) + \frac{4}{a^2}\sin^2\left( \frac{a k_1}{2}\right) + m^2\right]^{-1} \nonumber \\ 
\end{align}
If we define the Fourier component of the field as
\begin{align}
  \hat{\phi}(k,\tau) = \frac{a^2}{2\pi} \sum_{x \in \Gamma} e^{-i (k_0x_0+k_1x_1)}~\phi(x,\tau),  
\end{align}
we can then write the Fourier spectrum of the two-point correlation
function, according to the discretized-time version of
Eq.~\eqref{eq:def-correlations}, as the following dynamical average:
\begin{align}
  G(k) & = \langle \hat{\phi}^*(k) \hat{\phi}(k) \rangle =\frac{1}{\Delta\tau} \sum_{i=0}^M \phi^*(k,\tau_0+\delta\tau_i)\phi(k,\tau_0+\delta\tau_i),
\end{align}
where $\delta\tau_i = i\cdot \delta\tau$. For the Euclidean lattice we
have studied the lattice dynamics with the parameters and initial
conditions given at the end of Sec.~\ref{sec:simulation}, considering,
in addition, value of mass $m=3.0$ and nonlinearity coefficient
$\lambda=0.001$: the numerical value of the propagator in Fourier
space perfectly reproduces the expected dumbbell shape, as shown in
Fig.~[\ref{fig:euclid-propagator}].
\begin{figure}
\centering
  \includegraphics[width=0.7\columnwidth]{propagator_k_space_euclid-crop.png}
  \caption{{\it Real part of the two-point correlation function
      Fourier spectrum (Euclidean propagator) for a $\lambda\phi^4$
      theory in $d=2$ euclidean dimensions. Numerical value from the
      interacting theory with nonlinearity $\lambda=0.001$, lattice
      spacing $a=1.0$, lattice side $M_\mu=128$, mass $m=3.0$.}}
  \label{fig:euclid-propagator}
\end{figure}
We have also checked that the two-point correlation function exhibits
in real space the typical exponential decay $C(x) \sim e^{-m x}$.

\section{Minkowski lattice: linear and non-linear theory}
\label{sec:stability}

The numerical and analytical study of the free field theory in $1+1$
Minkowski space-time presents a new problem with respect to the
Euclidean space: the dynamics of quantum fluctuations for the {\it
  linear} non-interacting theory in the symplectic quantization
approach turns out to be {\it unstable}. This can be recognized
immediately from the free field equations in the continuum.\\
In the case of a purely quadratic potential $V[\phi] = \frac{1}{2} m^2
\phi^2$, the explicit solution of Eq.~(\ref{eq:eq-motion}) can be
obtained by exploiting the translational symmetry of space-time, which
allow to Fourier transform the equations:
\begin{align}
  \ddot{\phi}(k,\tau) + \omega_k^2~\phi(k,\tau) = 0,
\label{eq:harm-oscillator}
\end{align}
with
\begin{align}
\omega_k^2 = |{\bf k}|^2 +  m^2 - k_0^2. 
\end{align}
The general solution of Eq.~(\ref{eq:harm-oscillator}) can be then
written in terms of the initial conditions as
\begin{align}
  \phi(k,\tau) & = \phi(k,0) \cos(\omega_k\tau) + \frac{\dot{\phi}(k,0)}{\omega_k} \sin(\omega_k\tau)~~~~\forall~~\omega_k^2>0 \nonumber \\
  \phi(k,\tau) & = \phi(k,0) \cosh(z_k\tau) + \frac{\dot{\phi}(k,0)}{z_k} \sinh(z_k\tau)~~\forall~~\omega_k^2<0, 
\end{align}
where
\begin{align}
i z_k = \sqrt{\omega_k^2}.
\end{align}
Without any loss of generality and consistently with what we have done
numerically on the lattice, one can consider the following initial conditions:
\begin{align}
  \phi(k,0) &= 0 \nonumber \\
  \dot{\phi}(k,0) &= 1,
\end{align}
so that the general time-dependent solution reads as
\begin{align}
  \omega_k^2>0~~&\Longrightarrow~~\phi(k,\tau) = \frac{\sin(\omega_k\tau)}{\omega_k} \nonumber \\
  \omega_k^2<0~~&\Longrightarrow~~\phi(k,\tau) = \frac{\sinh(z_k\tau)}{z_k}.
\end{align}
Rewriting the generalized Hamiltonian in Fourier space we have
\begin{align}
  \mbH[\phi,\pi] = \frac{1}{2} \int d^dk~ \left(|\pi(k)|^2 + \omega_k^2 ~|\phi(k)|^2 \right),
  \label{eq:symp-Hamiltonian}
\end{align}
so that, by plugging into it the time-dependent solutions we have:
\begin{align}
  &\omega_k^2>0~~\Longrightarrow \mbH[\phi(\tau),\pi(\tau)] = \frac{1}{2} \int d^dk~\left[\cos^2(\omega_k\tau) + \sin^2(\omega_k\tau)\right] \nonumber \\\nonumber \\
  &\omega_k^2<0~~\Longrightarrow \mbH[\phi(\tau),\pi(\tau)] = \frac{1}{2} \int d^dk~\left[\cosh^2(z_k\tau) - \sinh^2(z_k\tau)\right]. 
\label{eq:compatibility-1}
\end{align}
Considering the expressions in Eq.~\eqref{eq:compatibility-1} we
realize that, despite the conservation of the symplectic quantization
Hamiltonian, it exists an infinite set of momenta, namely all $k$'s
with $\omega_k^2<0$, such that the {\it ``potential''} and {\it
  ``kinetic''} part of the generalized energy in
Eq.~\eqref{eq:symp-Hamiltonian}, namely $\mathbb{K}[\phi,\pi]$ and
$\mathbb{V}[\phi,\pi]$, both diverge exponentially with $\tau$. This
fact presents two problems, one conceptual and the second
numerical. The conceptual problem is represented by the fact that,
irrespectively to the behaviour of moments $\pi(k,\tau)$, which {\it
  might} also be regarded as unphysical auxiliary variables, we have
that also the contribution to generalized potential energy
$\mathbb{V}[\phi,\pi]$ (corresponding in practice to the relativistic
action) of an infinite amount of field modes $\phi(k,\tau)$ diverges
exponentially with $\tau$. This divergence of the field amplitude is
clearly unphysical: interpreting in fact as {\it ``particles''} the
modes of the free field, this would correspond to infinite growth of
the action of an isolated particle, which is clearly not observed in
the real world.
%%%
\begin{figure}[H]
  \centering
  \subfigure[{\it Behaviour of the normalized energy $(E(\tau)-E_0)/E_0$ vs $\tau$ for a scalar free theory ($\lambda=0$) with $m=1.0$, $a=1.0$, $M_\mu=128$, with initial conditions $\pi(k;0)=1$ and $\phi(k;0)=0$ for all $k$'s. Notice the exponential growth with $\tau$ which sets in after a short transient.}]{
    \includegraphics[width=0.45\columnwidth]{energy_div-crop.png}
    \label{fig:energy-free-periodic}
  }
  \hfill
  \subfigure[{\it Behaviour of the normalized energy $(E(\tau)-E_0)/E_0$ vs $\tau$ for a scalar theory with a small self-interaction term, $\lambda = 0.001$, and with $m=1.0$, $a=1.0$, $M_\mu=128$, with initial conditions $\pi(k;0)=1$ and $\phi(k;0)=0$ for all $k$'s. Oscillations are of order $\delta t$.}]{
    \includegraphics[width=0.45\columnwidth]{energy_cons-crop.png}
    \label{fig:energy-inter-periodic}
  }
  \caption{Comparison of normalized energy behaviour for scalar free theory and scalar theory with self-interaction.}
  \label{fig:energy-comparison}
\end{figure}
At the same time, attempting to study numerically the free-field
dynamics on Minkowski lattice, the leap-frog algorithm, which
proceedes alternating the update of kinetic and potential energy,
cannot handle the situation where the overall energy is conserved but
the two contributions diverge. Eventually, due to the accumulation of
numerical errors, total energy starts to diverge exponentially as well
with elapsing time, see Fig.~\ref{fig:energy-free-periodic} and the
discussion below.\\\\
Since in the present section we are just interested in the stability
of the theory, irrespectively of a realistic study of signals
propagation across the lattice, we have considered for simplicity
periodic boundary conditions. Let us remark that these conditions
would not be appropriate for a more realistic study of two-point
correlation functions with Minkowski metric, since in this case we
would like to probe the causal structure of space-time: a signal
escaping from the lattice at $+ct$ cannot appear back at $-ct$. For a
similar reason even fixed boundary conditions would not be
appropriate.\\\\
Using periodic boundary conditions we have checked numerically that
the symplectic quantization dynamics of a free scalar field suffers
from the pathology which can be conjectured already from the exact
solution: after a certain time the whole energy starts to grow
exponentially with $\tau$. In Fig.~[\ref{fig:energy-free-periodic}] we
present the results of simulations of the free-field with Minkowski
metric, all the parameters declared at the end of
Sec.~\ref{sec:simulation} and $m=1.0$, showing a clear evidence of the
exponential divergence with $\tau$. What seemed a good solution to
both the conceptual and numerical shortcomings of the free theory has
been to consider that the physically relevant theory is only the
interacting one: physical fields are always in interactions and the
``free-field theory'' is just an approximation, with some internal
inconsistencies which are revealed by the symplectic quantization
approach. Let us for instance consider a potential of the kind
\begin{align}
V[\phi] = \frac{1}{2} m^2 \phi^2 + \frac{1}{4} \lambda \phi^4,
\end{align}
for which the equations of motions in the continuum read as
\begin{align}
  & \ddot{\phi}(x,\tau)  =  - \partial_0^2\phi(x,\tau) + \sum_{i=1}^{d} \partial_i^2\phi(x,\tau) 
   - m \phi(x,\tau) - \lambda \phi^3(x,\tau).
  \label{eq:eq-motion-lambda}
\end{align}
Clearly, due to the non-linear term in
Eq.~\eqref{eq:eq-motion-lambda}, it is not possible anymore to
diagonalize the equations in Fourier space, so that both the sin/cos
and the sinh/cosh solutions cannot be taken into account as a
reference. Yet to be proven mathematically, the stability of
Eq.~\eqref{eq:eq-motion-lambda} is a quite delicate problem, since in
general for many Fourier components the equations are linearly
unstable. The intuition suggests that for each point of space-time $x$
the cubic force acts as a restoring term which prevents the amplitude
$\phi(x,\tau)$ to grow without bounds. This intuition has been
confirmed, up to the accuracy of our analysis, from our numerical
results. By using periodic boundary conditions, the parameters and
initial conditions declared at the end of Sec.~\ref{sec:simulation},
setting the non-linearity coefficient $\lambda=0.001$ we find that the
energy is no more divergent. The system relaxes to a stationary state
with oscillations of order $|E(t)-E_0|/E_0 = \mO(\delta \tau)$, as is
shown in Fig.~\ref{fig:corr-Mink-Fourier-space}. Let us
  remark, anticipating some of the upcoming results, that the
  instability of the deterministic dynamics of
  Eq.~\eqref{eq:sq-Ham-eq} in the case of a free theory is perfectly
  consistent with the shape of the canonical probability density
  $P[\phi] \sim \exp[S[\phi]/\hbar]$ for which we proved equivalence
  with the microcanonical weight $\rho_{\text{micro}}[\phi(x)] \sim
  \delta\left(\mA-\mathbb{H}[\phi,\pi]\right)$, as will be shown in
  Sec.~\ref{sec:pertequiv}: $P[\phi]$ turns out to be ill-defined for
  the action of a free scalar field.\\ \\

Having assessed the stability of the symplectic quantization dynamics
in the presence of non-linear interactions and periodic boundary
conditions, it is now time to consider, keeping the non-linearity
switched on, the more physical case of fringe boundary
conditions~\cite{NNH99}. This procedure will allow us to sample
numerically the Feynman propagator for small non-linearity, as will be
discussed in the next section.

\begin{figure}
  \includegraphics[width=0.50\columnwidth]{energy_k1.png}
  \includegraphics[width=0.50\columnwidth]{energy_k2.png}
  \caption{Behaviour of the time averaged harmonic
    $\overline{E}_{\text{harm}}(k,\tau)$ and kinetic
    $\overline{E}_{\text{kin}}(k,\tau)$ energies for two different
    choices of $k$, corresponding respectively to small (top panel)
    and large (bottom panel) scales. Non-linearity coefficient is
    $\lambda = 0.001$ and lattice parameters are with $m=3.0$,
    $a=1.0$, $M_\mu=128$, with initial conditions $\pi(k;0)=1$ and
    $\phi(k;0)=0$ for all $k$'s. For this choice of parameters there
    are no unstable modes, i.e. for all $k$'s we have $\omega_k^2>0$.}
  \label{fig:equipartition}
\end{figure}
\begin{figure}
  \includegraphics[width=0.5\columnwidth]{th_propagator.png}
  \includegraphics[width=0.5\columnwidth]{propagator_k_space.png}
  \caption{{\it Real part of the two-point correlation function
      Fourier spectrum $G(k_0,k_1) = \langle \phi^*(k_0,k_1)
      \phi(k_0,k_1)\rangle $ (Feynman propagator) for a
      $\lambda\phi^4$ theory in $1+1$ space-time dimensions. {\it
        Top}: theoretical value of the free propagator with lattice
      spacing $a=1.0$, lattice side $M_\mu=128$, mass $m=3.0$; {\it
        Bottom}: numerical value from the interacting theory with the
      same parameters and nonlinearity $\lambda=0.001$. Initial
      conditions are set to $\phi(k;0)=0$ and $\pi(k;0)=1$ for all
      $k$'s. For this choice of parameters there are no unstable
      modes, i.e. for all $k$'s we have $\omega_k^2>0$.}}
  \label{fig:corr-Mink-Fourier-space}
\end{figure}
%%
%%
%%%%%%%%%%%%%%%%%%%%%%%%%%%%%%%%%%%%%%%%%%%%%%%%%%%%%%%%%%%%%%
\section{Feynman propagator: numerical results}
\label{sec:Feynman}
In the previous section we have shown how the presence of non-linear interactions solves the instability problem of the linear theory, still keeping periodic boundary conditions. But periodic boundary conditions are clearly unphysical, because one of the directions of our lattice corresponds to $ct$, so that periodicity of the boundaries is clearly meaningless. We need to devise a strategy to mimic the free propagation of any kind of signal across the boundaries as if outside there was an infinitely large lattice. This strategy is provided by the use of fringe boundary conditions, introduced in Sec.~\ref{sec:simulation}.\\
In this part of the paper we will therefore provide the numerical evidence that for perturbative values of the non-linearity coefficient $\lambda$ we recover qualitatively the correct shape of the free Feynman propagator.\\
The strategy is very simple: having set the coefficient of the non-linear interaction $\lambda$ to a small but finite value, $\lambda=0.001$, we have run the symplectic dynamics with fringe boundary conditions until stationarity is reached at a certain time, which we call $\tau_{\text{eq}}$. According to the premises of Sec.~\ref{sec:SQ-foundations}, where we assumed that at long enough times the symplectic quantization dynamics allows us to sample an equilibrium ensemble, we have checked that equipartition between positional and kinetic degrees of freedom is in fact reached.  In Fig.~\ref{fig:equipartition} is shown how, for two given choices of $k=\lbrace k_0,k_1\rbrace$ (corresponding respectively to small and large scales), we have that $\overline{E}_{\text{harm}}(k,\tau)$ and $\overline{E}_{\text{kin}}(k,\tau)$ reach asymptotically a value close to $1/2$, starting respectively from $E_{\text{harm}}(k,0)=0$ and
$E_{\text{kin}}(k,0)=1$, where the two energies are defined
respectively as
\begin{align}
  \overline{E}_{\text{harm}}(k,\tau) &= \frac{1}{\tau} \int_0^\tau ds~\frac{1}{2}\omega_k^2 |\phi(k,s)|^2 \nonumber \\
  \overline{E}_{\text{kin}}(k,\tau) &= \frac{1}{\tau} \int_0^\tau ds~\frac{1}{2}|\pi(k,s)|^2.
\end{align}
We have found that this standard equipartition condition is fulfilled well when all $k$'s in the lattice are such that $\omega_k^2>0$, while the stationary state reached when a finite fraction of the modes is such that $\omega_k^2<0$ has less trivial properties, which will be analysed in further details elsewhere.\\
Having thus assessed that the system reaches some equilibrium/stationary state within some time $\tau_{\text{eq}}$, we have computed for all times $\tau > \tau_{\text{eq}}$ the Fourier spectrum of the two-point correlation function $G(k) = \langle \phi^*(k) \phi(k)\rangle$ by averaging (quantum) fluctuations over intrinsic time. That is, we have defined an interval $\Delta \tau$ large enough and we have computed
\begin{align}
\langle \phi^*(k) \phi(k) \rangle = \frac{1}{\Delta \tau} \sum_{i=0}^M
\phi^*(k,\tau_{\text{eq}} + \tau_i) \phi(k, \tau_{\text{eq}} + \tau_i),
\end{align}
where $\tau_i = i\cdot\delta\tau$ and $\Delta\tau = M \delta\tau$.\\
In Fig.~\ref{fig:corr-Mink-Fourier-space} we show (bottom panel) the result for the Fourier spectrum of the two-point correlation function obtained by setting all the parameters of the simulation and the initial conditions as declared at the end of Sec.~\ref{sec:simulation}, apart from the value of the mass that is set here at $m=3.0$ in order to better appreciate the shape of the propagator, and taking the value $\lambda=0.001$ for the non-linearity parameter. In order to compare our numerical data at small non-linearity with the theory, we have also reported in the top panel of Fig.\ref{fig:corr-Mink-Fourier-space} the theoretical shape of the free Feynman propagator $G_{\text{th}}(k_0,k_1)$ on a discretized space-time grid in $1+1$ dimensions, using for the lattice the same parameters of the simulation, i.e., $a=1.0$, $m=1.0$, and $M_\mu=128$, where $G_{\text{th}}(k_0,k_1)$ reads as
\begin{align}
G_{\text{th}}(k_0,k_1) = \left[ \frac{4}{a^2} \sin^2\left( \frac{a k_0}{2} \right) - \frac{4}{a^2} \sin^2\left( \frac{a k_1}{2}\right) -m^2 \right]^{-1}.
\end{align}
\begin{figure}
  \includegraphics[width=0.5\columnwidth]{correlationfunction_x_fringe.png}
  \includegraphics[width=0.5\columnwidth]{correlationfunction_t_fringe.png}
  \caption{{\it Real space two-point correlation function for a
      $\lambda\phi^4$ theory in $1+1$ space-time dimensions with
      fringe boundary conditions, lattice spacing $a=1.0$, lattice
      side $M_\mu=128$, mass $m=1.0$ and nonlinearity
      $\lambda=0.001$. {\it Top}: exponential decay along the
      direction parallel to the $x_1$ axis; {\it bottom}:
      oscillations along the direction parallel to the $x_0=ct$ axis.}}
  \label{fig:corr-Mink-real-space}
\end{figure}
Let us stress the beautiful qualitative agreement between the theoretical prediction of the free propagator and the numerical results: at variance with the Euclidean propagator, which is a function decreasing monotonically in all directions moving away from the origin (see Fig.~\ref{fig:euclid-propagator} above), we find that the Feynman propagator sampled numerically here has the characteristic shape of a saddle, denoting a different behaviour between {\it time-like} directions and {\it space-like} directions. This is the first and incontrovertible strong evidence that the symplectic quantization approach opens up new possibilities so far out of reach within the Euclidean formulation of lattice field theory. Even more clear is the signature of the causal structure of space-time probed by means of the new approach if we look at the two-point correlation function in real space. According to the theoretical predictions for the free theory in the continuum one would expect undamped oscillations along the purely {\it time-like} directions and an exponential decay along the purely {\it space-like} directions for the Feynman propagator $\Delta_F(x-y)$:
\begin{align}
\Delta_F (x-y) = \frac{1}{(2\pi)^2} \int d^2k~\frac{e^{ik(x-y)}}{k^2-m^2},
\end{align}
with
\begin{align}
\Delta_F (x-y)  &\sim e^{im |x-y|} \quad~\text{for}~~~ x-y ~|| ~x_0 \nonumber \\    
\Delta_F (x-y)  &\sim e^{-m |x-y|} \quad\text{for}~~~  x-y ~|| ~x_1. 
\end{align}
Clearly, when the same correlation function is sampled on a finite and discrete grid there will be finite-size effects at play so that, for instance, also the oscillations along the time-like direction will be slightly modulated by a tiny exponential decay: this is precisely what we find in numerical simulations. In Fig.~[\ref{fig:corr-Mink-real-space}] are shown, respectively in top and bottom panels, the exponential decay along the purely space-like direction and the oscillations along the purely time-like direction, obtained for the following choice of parameters: $a=1.0$, $M_\mu=128$, mass $m=1.0$ and nonlinearity $\lambda=0.001$. Let us notice that the value of the mass which can be obtained from either the fit of the exponential decay as $C(\Delta x_1) \sim e^{- m \Delta x_1}$ or the oscillating part as $C(\Delta x_0) \sim e^{ i m \Delta x_0}$ is $m \sim 2.06\pm 0.04$, i.e., quite different from the value $m=1$ put in the Lagrangian. This effect, which we do not find for the deterministic dynamics in Euclidean space-time, is most probably a finite-size effect related to the propagation of signals across fringe boundary conditions. We made some attempts, discussed in Appendix~\ref{appendix:fringe}, to investigate a possible interplay between the measured mass discrepancy with the way fringe boundary conditions are imposed, but we didn't find any clear indication on the possible origin of the effect. A stronger effort is for sure necessary to put under control the finite-size effects related to fringe boundary conditions: we plan to devote another paper to this problem.
%%%%%%%%%%%%%%%%%%%%%%%%%%%%%%%%%%%%%%%%%%%%
\section{Canonical form of Minkowskian statistical mechanics}
\label{sec:pertequiv}

In Sec.~\ref{sec:Feynman} we have shown that the Hamiltonian dynamics
of a quantum field theory with an additional time paramter $\tau$ and
corresponding conjugated momenta allows to recover qualitatively well
the shape of the free Feynman propagator for a small value of the
interaction constant $\lambda$.  It is therefore legitimate to wonder
which is the precise relation between the correlation functions
obtained in this generalized microncanonical ensemble and the one
generated by the Feynman path integral and/or the corresponding
Euclidean Field Theory. As a first step in this direction we propose
an explicit calculation of the microcanonical partition function in
the large-$M$ limit, where $M$ is the number of degrees of
freedom. The calculation shows that also for this peculiar system,
where the microcanonical ensemble is built on the conservation of an
action rather than an energy functional, the sampling of fluctuations
in this ensemble is formally equivalent to the sampling of a {\it
  canonical} one at a {\it temperature} $\hbar$, namely fields
fluctuations are sampled with probability $\exp(S/\hbar)$.\\
%%
%From the shape of the theory it turns also out to be evident that the shape of two-point correlation functions, at least for small values of the interaction constant $\lambda$ can be reasonable expected to be qualitatively similar to standard two point correlation functions, whereas the analytic continuation to standard field theoretical propagators for non-perturbative values of $\lambda$, though very promising, deserves to be studied more in detail. 
%%
The explicit computation of the microcanonical partition function in the large-$M$ limit proceeds then as follows. As it is customary for the purpose of computing correlation functions, we assume the presence of an external source $J(x)$ linearly coupled to the field:
\begin{align}
  \Omega[\mA,J] = \int \mD\phi\mD\pi~\delta\left(\mA-\mathbb{H}[\phi,\pi]+\int d^dx\,J(x) \phi(x)\right).
  \label{eq:gen-micro}
\end{align}
Since we are using the lattice a regularizer for the theory, the field can be conveniently expanded in an orthonormal basis as follows~\cite{STROMINGER}:
\begin{equation}
	\phi(x) = \sum_{n=1}^M \phi_n(x) c_n,
\end{equation}
where
\begin{equation}
	\int d^dx\, \phi_n(x) \phi_m(x) = \delta_{mn}.
\end{equation}
We can then define a {\it finite} measure over the field configuration, reading as:
\begin{equation}
	\int \mathcal{D}_M \phi \equiv \prod_{n=1}^M \int_{-\infty}^{\infty} dc_n.
\end{equation}
Contrary to the usual convention, there is no $\hbar$ in this measure. In a $d$-dimensional box of volume $L^d$ with lattice spacing $a$, the number of basis functions is~\cite{STROMINGER}:
\begin{equation}\label{eq:Mlambda}
	M = \frac{L^d}{a^d} = \frac{1}{\pi^d} L^d \Lambda^d,
\end{equation}
where $\Lambda = \pi/a$ is the momentum cutoff. Let us specify that $M$ is not the number of classical dynamical degrees of freedom, which grows on the contrary simply as $(L\Lambda)^3$. From Eq.~\eqref{eq:Mlambda} we see that the {\it field} limit $M\rightarrow\infty$ can be obtained either as the continuum limit, $\Lambda \rightarrow \infty$, or as the thermodynamic limit, $L \rightarrow \infty$. Nevertheless, we will make a crucial step in the calculation of the microcanonical partition function in the large-$M$ which is well justified {\it only} in the {\it continuum} limit and not in the thermodynamic one, so that from here on we will refer to the limit $M\rightarrow\infty$ as the continuum limit.\\\\ 
By lightening the notation according to the following conventions
\begin{align}
	\pi^2 &\equiv  \int d^dx\, \pi^2(x)\nonumber\\
	J \cdot \phi &\equiv \int d^dx\, J(x) \phi(x).
\end{align}
we can then rewrite the partition function on the lattice as:
\begin{equation}
	\Omega[\mA,J] = \int \mathcal{D}\phi_M\mathcal{D}\pi_M\, \delta\left(\mA-\frac{\pi^2}{2} +S[\phi]+J\phi\right)\,.
\end{equation}
The functional integration over $\pi(x)$ can be then done by taking advantage of the following formula, valid for $R^2>0$:
\begin{align}
  I_M(R) &= \int_{-\infty}^\infty dx_1\ldots dx_M ~\delta\left( \frac{1}{2}\sum_{i=1}^M x_i^2 - R^2 \right) = \frac{(2\pi)^{\frac{M}{2}}}{\Gamma\left(\frac{M}{2}\right)}~R^{M-2}, 
\end{align}
from which, putting $R = (\mA+S[\phi]+J\phi)^{\frac{1}{2}}$, we get:
\begin{align}
  \label{eq:omega-step}
  \Omega[\mA,J] = \frac{(2\pi)^{\frac{M}{2}}}{\Gamma\left(\frac{M}{2}\right)} \int \mathcal{D}\phi_M\left(\mA+S[\phi]+J\phi\right)^{\frac{M}{2}-1}.
\end{align}
The positivity of $R^2=\mA+S[\phi]+J\phi$, which is crucial for the whole calculation, is ensured by construction of the microcanonical ensemble, since the kinetic energy term related to conjugate momenta is positive definite. In order to consider a large-$M$ limit in the computation is convenient at this stage to rewrite the partition function in Eq.~\eqref{eq:omega-step} in the following form, which puts in evidence the dependence on $M$:
\begin{align}
	\Omega[\mA,J] = \kappa_M\int \mathcal{D}\phi_M\,\exp\left\{\left(\frac{M}{2}-1\right)\ln\left(\mA+S[\phi]+J\phi\right)\right\}, \nonumber \\
    \label{eq:Omega-less-momenta}
\end{align}
where $\kappa_M= (2\pi)^{\frac{M}{2}}/\Gamma\left(M/2\right)$. In order to now fulfill the same quantization constraint used for the numerical simulation discussed in the previous section we assign $\hbar$ to every degree of freedom, equally sharing this amount among "positional" and "kinetic" components. \\
Since momenta have been integrated out, to make the integral finite, in expression Eq.~\eqref{eq:Omega-less-momenta} we need to fix $\mA$ to half of the total value, since we need to account only for "momenta" degrees of freedom, namely we write
\begin{align}
\mA_z = \frac{\hbar M}{2z}\,,
\end{align}
while we consider the counterterms for the "positional" degrees of freedom to be already in the action.
We have introduced at this point the dimensionless parameter $z$ in order to be able to tune the value of the average {\it quantum of action} per degree of freedom in the final expression that we will derive for $\Omega[\mA,J]$ and also with the purpose to highlight how the present theory connects to ordinary Feynman path integral by analytic continuation in $z$. We now proceed to expand the partition function $\Omega[\mA_z,J]$ in powers of $J$ so that we can write explicitly the generating functional in terms of correlators. By doing this, ignoring the subleading $O(1)$ in $M$ term in the exponent, we get:

\begin{align}
	\Omega[\mA_z,J] =& ~\kappa_M \left(\frac{\hbar M}{2z}\right)^{\frac{M}{2}} \sum_{n=0}^{\infty}\frac{1}{n!}\left(\frac{z}{\hbar}\right)^n\left(\frac{2}{M}\right)^n\frac{\Gamma(\frac{M}{2}+1)}{\Gamma(\frac{M}{2}+1-n)}\nonumber\\
	&\int d^dx_1\ldots d^dx_n~J(x_1)\ldots J(x_n) \int \mathcal{D}_M\phi~\phi(x_1)\ldots \phi(x_n) \left(1+\frac{2}{M}\frac{z}{\hbar}S[\phi]\right)^{\frac{M}{2}-n}\,.
\end{align}
Now, we proceed to expand $\left(1+\frac{2}{M}\frac{z}{\hbar}S[\phi]\right)^{\frac{M}{2}-n}$ in powers of $1/M$. A tedious but straightforward calculation yields:
\begin{equation}
  \left(1+\frac{2}{M}\frac{z}{\hbar}S[\phi]\right)^{\frac{M}{2}-n} =
  e^{\frac{z}{\hbar}S[\phi]} \exp\Bigg(\sum_{j=1}^{\infty}(-1)^j~\frac{2(2j-2)!!}{(j+1)!}\left(\frac{z}{\hbar}\frac{S[\phi]}{M}\right)^j~[j S[\phi]+(j+1)n]\Bigg)\,.
\end{equation}
from which we have
\begin{align}
%\label{eq:latticeomega}
	&\Omega[\mA_z,J] = \kappa_M~\left(\frac{\hbar M}{2z}\right)^{\frac{M}{2}}~\sum_{n=0}^{\infty}\frac{1}{n!}\left(\frac{z}{\hbar}\right)^n\left(\frac{2}{M}\right)^n\frac{\Gamma(\frac{M}{2}+1)}{\Gamma(\frac{M}{2}+1-n)} \nonumber\\
	&\int d^dx_1\ldots d^dx_n~J(x_1)\ldots J(x_n)~\left\langle \phi(x_1)\ldots \phi(x_n)~e^{\sum_{j=1}^{\infty}(-1)^j~\frac{2(2j-2)!!}{(j+1)!}\left(\frac{z}{\hbar}\frac{S[\phi]}{M}\right)^j~[j S[\phi]+(j+1)n]}\right\rangle,
\end{align}
where the expectation value, denoted $\langle\,\cdot\,\rangle$, is taken with respect to the weight $\exp\left(z S[\phi]/\hbar\right)$ with $S[\phi]$ being the renormalized action. Consider now the correlators:
\begin{align}
  &\left\langle \phi(x_1) \ldots \phi(x_n)~e^{\sum_{j=1}^{\infty}(-1)^j~\frac{2(2j-2)!!}{(j+1)!}\left(\frac{z}{\hbar}\frac{S[\phi]}{M}\right)^j~[j S[\phi]+(j+1)n]}\right\rangle=\nonumber
  \\&=\left\langle \phi(x_1)\ldots \phi(x_n) \right\rangle+\sum_{j=1}^{\infty} \frac{c_j(M,n)}{M^j} \left\langle \phi_1(x_1)\ldots\phi_n(x_n)~S^{j}[\phi]\right\rangle,
  \label{eq:latticeomega2}
\end{align}
where in the second line we performed a large-$M$ expansion of the exponential. It turns out that the coefficients $c_j(M,n)$ are polynomials in $M$ with an asymptotic behavior of the kind 
\begin{equation}
    \frac{c_j(M,n)}{M^j} = o\left(\frac{1}{M}\right).
\end{equation}
We need now to ascertain whether $\Omega[\mA_z,J]$ has a sensible field limit, $M\rightarrow\infty$. We begin by noticing that 
\begin{equation}
\lim_{M\to\infty}\left(\frac{2}{M}\right)^n \frac{\Gamma(\frac{M}{2}+1)}{\Gamma(\frac{M}{2}+1-n)}\to 1,
\end{equation}
which tells us that the coefficient of each term in the sum goes to unity in the large-$M$ limit. We make now the (very reasonable) assumption that all the insertions of powers of the renormalized action $c_j(M,n) S[\phi]^{j}/M^j$ in the correlators appearing in Eq.~\eqref{eq:latticeomega2} go smoothly to zero in the continuum limit: this is in fact equivalent to assume that in the continuum limit the renormalized action remains finite in a finite volume, i.e., $S[\phi]/M \rightarrow 0$ when $M\rightarrow\infty$. Clearly the assumption that $S[\phi]$ remains finite in the limit $M\rightarrow\infty$ would not equally apply to thermodynamic limit, where we expect the renormalized action to be extensive, $S[\phi] \sim M$. We therefore have that for each term of the series the following holds:
\begin{equation}
\lim_{M\to\infty}\left\langle \phi_1(x_1)\ldots,\phi_n(x_n) \frac{c_j(M,n)}{M^j}S^{j}[\phi]\right\rangle = 0 ~~~\forall ~ j,
\label{eq:null-insertion}
\end{equation}
which finally leads to:

\begin{align}
&\Omega[\mA_z,J] =\nonumber\\
&= \kappa_M~\left(\frac{ \hbar M}{2z}\right)^{\frac{M}{2}}~\sum_{n=0}^{\infty}\frac{1}{n!}\left(\frac{z}{\hbar}\right)^n
~\int d^dx_1\ldots d^dx_n~J(x_1)\ldots J(x_n)~\left\langle \phi(x_1)\ldots \phi(x_n)\right\rangle  \nonumber \\
& = \kappa_M~\left(\frac{ \hbar M}{2z}\right)^{\frac{M}{2}} \int \mathcal{D}\phi_M~e^{\frac{z}{\hbar}S[\phi]}\sum_{n=0}^{\infty}\frac{1}{n!}\left(\frac{z}{\hbar}\right)^n
\int d^dx_1\ldots d^dx_n\, J(x_1)\ldots J(x_n)\phi(x_1)\ldots \phi(x_n) \nonumber \\
& = \kappa_M~\left(\frac{ \hbar M}{2z}\right)^{\frac{M}{2}} \int \mathcal{D}\phi_M~e^{\frac{z}{\hbar}S[\phi]+\frac{z}{\hbar} \int d^{d}x~J(x)\phi(x)}\,.
\label{eq:ultimate-result}
\end{align}
From Eq.~\eqref{eq:ultimate-result} we can therefore conclude that, up to irrelevant multiplicative constants, the symplectic quantization microcanonical generating functional in then {\it continuum} limit takes the form:
\begin{align}
	\Omega[\hbar/z,J]=\int \mathcal{D}\phi\,\exp\left(\frac{z}{\hbar}S[\phi]+\frac{z}{\hbar}J\phi\right).
    \label{eq:final-Omega}
\end{align}
The choice of $z$ corresponding to the simulations presented in the first part of this work is $z=1$: in this case the expression of $\Omega[\hbar,J]$ obtained in Eq.~\eqref{eq:final-Omega} tells us that the correlation functions measured the Hamiltonian dynamics of symplectic quantization are identical, to the leading order in $M$ and provided that ergodicity holds, to those obtained from a canonical probability distribution of the kind
\begin{align}
P[\phi] = \frac{e^{S[\phi]/\hbar}}{\Omega[\hbar]}.
\label{eq:final-P}
\end{align}
It comes quite natural at this point a short remark on how the
  continuum limit can be possibly be considered in our numerical
  setup, in order to gain full consistency between simulation results
  and the present analytical derivation. Very simply we assume that in
  the symplectic quantization framework the continuum limit can be
  taken exactly as in ordinary lattice quantum field theory. One has
  to consider the limit of a vanishing lattice spacing, $a\rightarrow
  0$, while tuning the bare parameters so as to hold a chosen
  renormalization group (RG) invariant observable fixed. In practice,
  the prescription to consider the continuum limit can be realized as
  follows: after having identified a given RG-invariant quantity $X$,
  one has to perform simulations at several values of the lattice
  spacings $a$, then adjusting the coupling(s) and the mass(es) so
  that the RG-invariant observable $X$ remains constant, thus ensuring
  a correct RG flow. The results of simulations can be then used to
  extrapolate smoothly the values of physical observables at $a=0$,
  i.e., in the continuum limit. Let us also notice that this
  procedure, while being necessary for any future use of the
  Symplectic Quantization approach to extract physical information for
  realistic theories, it is not particularly interesting for an
  asymptotically trivial theory such as $\lambda \phi^4$, where the
  RG-flow is to a Gaussian fixed point, which makes the continuum limit
  trivial.

Let us conclude with two main remarks about the results in
Eqns.~\eqref{eq:final-Omega}, \eqref{eq:final-P}. First of all we have
shown that the microcanonical sampling is equivalent to the sampling
from a probability distribution $P[\phi]$ which is well defined for an
interacting theory with a potential bounded from below, since for
configuration of the field with large values and smooth variations we
have approximatively

\begin{align}
    e^{S[\phi]/\hbar} \sim e^{-V[\phi]/\hbar}.
\end{align}

This is completely in agreement with the results of numerical
simulations in the microcanonical ensemble, where the Hamiltonian
dynamics of the free theory develops run-away solutions. Second, the
result of our derivation in Eq.~\eqref{eq:final-Omega} shows us that
this new ``canonical Minkowskian measure'' can be connected to the
standard Feynman path integral by means of analytic continuation in
the dimensionless parameter $z$. More investigations in this direction
are actually in progress.
%%%%%%%%%%%%%%%%%%%%%%%%%%%%%%%%%%%%%%%%%%%%%%%%%%%%%%%%%%%%%%%%%%%%%%%%%%
\section{Conclusions and Perspectives}
\label{sec:conclusions}
In this work we have presented the first numerical test of symplectic
quantization, a new functional approach to quantum field
theory~\cite{GL21,G21} which allows for an importance sampling
procedure directly in Minkowski space-time. The whole idea, which
parallels the one of stochastic quantization, is based on the
assumption that fields have a dependence of an additional time
parameter, the intrinsic time $\tau$, with respect to which conjugated
momenta $\pi(x)$ are defined. Quantum fluctuations of the fields are
sampled by means of a deterministic dynamics flowing along the new
time $\tau$, which controls the internal dynamics of the system and is
distinguished from the coordinate time of observers and clocks.  Such
a dynamics is generated by a generalized Hamiltonian where the
original relativistic action plays the role of a potential energy part
and therefore fluctuates naturally along the flow of $\tau$. This
whole construction does not need any sort of rotation from real to
imaginary time to be consistent and to efficiently allow the
numerical sampling of field fluctuations. Furthermore, under the
hypothesis of ergodicity, symplectic quantization allows to define a
generalized microcanonical ensemble which represents a
probabilistically well defined functional approach to quantum field theory. In Sec.~\ref{sec:pertequiv} we have shown that the microcanonical partition function corresponding to the symplectic quantization dynamics is simply connected by means of an integral transformation to the Feynman path integral, thus implying that also all the disconnected correlation functions measured from the symplectic quantization approach are connected by means of an integral transformation to quantum field theoretic correlations. This said, there is also another possible way to connect the microcanonical functionals studied in this work to the standard Feynman path integral. We have shown that it is possible to explicitly compute $\Omega[\hbar/z,J]$ in the continuum limit, where it turns out to be equivalent to a Minkowskian statistical mechanics theory with canonical weight $P[\phi] \propto \exp(zS[\phi]/\hbar)$. First of all, consistently with the results of our simulations, it must be noticed that this canonical probability is well defined for a Minkwoskian theory, provided that the potential is bounded from below, it is therefore a very promising tool to study non-perturbative problems in causal space-time. Second, the above canonical expression suggests that a punctual correspondence with standard quantum field theory can be drawn by analytically continuing along a suitable integration path the above weight in the complex $z$ plane \cite{Witten:2010cx,witten2010newlookpathintegral}. This last path looks very promising and is currently under investigation.

%among the main result of this paper is the proof that the
%multipoint correlation functions generated by this generalized
%microcanonical ensemble are identical to the correlation functions, at least for the case of a scalar field theory, to those generated by the standard Feynman path integral approach. Our analytical proof, which for didactical purposes we have presented accounting for quantum fluctuations both at the one-loop level and then in general,is backed by the numerical evidence that the shape of the free Feynman propagator can be efficiently sampled for a $\lambda \phi^4$ real theory for a small value of $\lambda$, making the overall proposal of the new approach to quantum field theory very robust. Among the short-term perspectives there is for sure, on the technical side, the realization of a more careful study of the finite-size effects induced by fringe boundary conditions and the test of the symplectic quantization approach in the presence of non-perturbative values of the nonlinearity $\lambda$. Theoretically, it is compelling to extend this new framework to include gauge fields and fermions, in order to test its power on real lattice QCD simulation. For the moment we have worked on an extension of the presented equivalence proof between symplectic quantization and standard QFT to the case of pure gauge fields, a task which is non-trivial but at the same, when the correct strategy is devised, leads to a straightforward calculation: this is presently work in progress~\cite{S24}.
%
\acknowledgments
We thank M. Bonvini, P. Di Cintio, R. Livi, S. Matarrese, F. Mercati,
M. Papinutto, A. Polosa, A. Ponno, L. Salasnich, G. Santoni,
N. Tantalo and F. Viola for valuable discussions on the subject of
this paper. In particular we want to thank M. Bonvini, F. Mercati and
L. Livi for several interactions at the beginning of this
project. G.G. thanks the Physics department of Sapienza for its kind
hospitality during several periods during the preparation of this
work. G.G. acknowledge partial support from the project MIUR-PRIN2022,
{\it ``Emergent Dynamical Patterns of Disordered Systems with
  Applications to Natural Communities''}, code 2022WPHMXK.
\newpage 
\appendix
%%%%%%%%%%%%%%%%%%%%%%%%%%%%%%%%%
\section{Numerical Algorithm}
\label{appendix:algorithm}
All numerical cacalculations in this paper have been performed using a
siplitting algorithm of second order, which takes advance of the
generalized Hamiltonian separability. Using the notation of
Sec.~\ref{sec:SQ-foundations}, the algorithm can be characterized as a
map
\begin{align}
\Psi_{\delta\tau}: \phi(x,\tau), \pi(x,\tau) ~~\longrightarrow~~ \phi(x,\tau+\delta\tau), \pi(x,\tau+\delta\tau), 
\end{align}
with the following structure
\begin{align}
\Psi_{\delta\tau} = \Phi^{\delta\tau/2}_{\mathbb{K}} \circ  \Phi^{\delta\tau}_{\mathbb{V}} \circ  \Phi^{\delta\tau/2}_{\mathbb{K}}, 
\end{align}
where $\Phi^{\delta\tau/2}_{\mathbb{K}}$ denotes the Hamiltonian flow
of $\mathbb{K}[\pi]$, i.e., the flow of generalized momenta, while
$\Phi^{\delta\tau}_{\mathbb{V}}$ denotes the Hamiltonian flow of
$\mathbb{V}[\phi]$, i.e., the flow of generalized coordinates (in this
case, the field). In formulae, each time step of the algorithm is
represented by the following sequence of operations, to be realized
for each point of $x$ of the lattice:
\begin{figure}[H]
  \centering
  \subfigure[Energy fluctuations $\delta E(\delta\tau)$ as a function of the timestep $\delta\tau$ of the numerical algorithm in the case of Minkowski metric and {\it periodic} boundary conditions. Energy conservation at the algorithmic precision, i.e., $\delta E(\delta\tau)\sim \delta\tau^2$, is fulfilled.]{
    \includegraphics[width=0.45\columnwidth]{energyscale_periodic.png}
    \label{fig:scaling-DE-periodic}
  }
  \hfill
  \subfigure[Energy fluctuations $\delta E(\delta\tau)$ as a function of the timestep $\delta\tau$ of the numerical algorithm in the case of Minkowski metric and {\it fringe} boundary conditions. Energy conservation at the algorithmic precision, i.e., $\delta E(\delta\tau)\sim \delta\tau^2$, is fulfilled.]{
    \includegraphics[width=0.45\columnwidth]{energyscale_fringe.png}
    \label{fig:scaling-DE-fringe}
  }
  \caption{Energy fluctuations $\delta E(\delta\tau)$ for different boundary conditions in the Minkowski metric.}
  \label{fig:scaling-DE-comparison}
\end{figure}

\begin{align}
  \pi(x,\tau+\delta\tau/2) &= \pi(x,\tau) + \frac{\delta\tau}{2} \cdot F[\phi(x,\tau)]~~~~~~~~\forall~x \nonumber \\
  \phi(x,\tau+\delta\tau) &= \phi(x,\tau) + \delta\tau \cdot \pi(x,\tau+\delta\tau/2) ~~~\forall~x \nonumber \\
  \pi(x,\tau+\delta\tau) &= \pi(x,\tau+\delta\tau/2) + \frac{\delta\tau}{2} \cdot F[\phi(x,\tau+\delta\tau)]~~~\forall~x \nonumber \\
\end{align}
The splitting algorithm which we have just described is usually known
as the leapfrog algorithm, the name coming from the fact the updated
of generalized positions and velocities takes place at interleaved
time points. Given $ E_0 = \mathbb{H}[\phi(x,0),\pi(x,0)] $ and
$E(\tau) = \mathbb{H}[\phi(x,\tau),\pi(x,\tau)]$, where $\phi(x,\tau)$
and $\pi(x,\tau)$ are the numerical solutions computed at $\tau$, the
leapfrog dynamics has the following algorithmic bound on energy
fluctuations
\begin{align}
  \delta E(\delta\tau) = \langle |E(\tau)/E_0-1| \rangle ~\propto ~ \delta\tau^2
  \label{eq:LF-bound}
\end{align}
We have verified that the bound in Eq.~\eqref{eq:LF-bound} is
fulfilled by the fluctuations of both the Hamiltonian $E(\tau) =
\mathbb{H}[\phi(x,\tau),\pi(x,\tau)]$ in the case of Minkowski metric
with periodic boundary conditions and the total Hamiltonian (system +
boundary layers) $\mathbb{H}_f[\phi(x,\tau),\pi(x,\tau)]$ in the case
of {\it fringe} boundary conditions (See
Eq.~\eqref{eq:fringe-Hamiltonian} and the following discussion for the
definition of $\mathbb{H}_f[\phi,\pi]$). In
Fig.\ref{fig:scaling-DE-periodic} and Fig.\ref{fig:scaling-DE-fringe}
is shown the behavior of $\delta E(\delta\tau)$ as a function of
$\delta\tau$ respectively for the case of periodic and fringe boundary
conditions.

\section{Fringe boundaries}
\label{appendix:fringe}

In this section we present results of a preliminary investigation into the sensitivity of the measured mass to the parameters governing the fringe boundary conditions. 
These numerical tests were conducted with the same parameters used for Fig.~\ref{fig:corr-Mink-real-space} in the main text, namely $m=1$, $\lambda=0.001$, $a=1$ and $L=128$, unless otherwise specified. Our findings are summarized in Table~\ref{tab:fringe_tests} and discussed below.\\
In the first place, we investigated the behavior of the measured mass upon varying the damping parameter $\varepsilon$, while keeping it constant across space. The top section of Table~\ref{tab:fringe_tests} shows that even by varying $\varepsilon$ over ten orders of magnitude (from $10^{-5}$ down to $10^{-15}$) we do not find evidence of any trend in the change of the measured mass, which therefore seems not affected by variations of $\varepsilon$. We also found that simply increasing the fringe region thickness $L_\text{fringe}$ from $L$ to $2L$ also makes apparently no difference.\\
We then investigated whether a spatially varying profile $\varepsilon(x)$ may influence the finite-size effects on the measured mass. We considered both a linear and an exponential decay for $\varepsilon(x)$ across the fringe layer. As shown in the middle section of Table~\ref{tab:fringe_tests}, a linear decay profile seems to have no effect on the measured mass. On the contrary, by exploiting an exponential decay of the form $\varepsilon(x)=\exp(-cx)$ we find an encouraging signal: the measured mass value seems to have a trend towards the expected value as long as the decay becomes sharper (i.e., for larger $c$).\\
Finally, we made a very preliminary investigation on how the measured mass value depends on the ratio $L_{\text{fringe}}/L$, upon increasing $L$ at fixed $L_{\text{fringe}}$. From the bottom section of Table~\ref{tab:fringe_tests} we have some preliminary indication that decreasing the ratio $L_{\text{fringe}}/L$ there is a trend towards the expected value of the mass, but a more systematic investigation is clearly in order.

%In summary, this investigation reveals several key points:
%\begin{enumerate}
 %   \item The measured mass has a non-trivial dependence on the implementation of the fringe boundaries.
  %  \item A sharp, exponential-like transition at the physical boundary is more effective than a slow, linear one.
   % \item The issue is coupled with finite-size effects, as larger lattice volumes improve the result.
%\end{enumerate}
%}

%While these tests confirm that the fringe conditions are indeed related to the mass issue, they also show that a simple parameter change is insufficient to fully resolve it. A complete solution would likely require a systematic, large-scale study of the combined effects of lattice volume, the ratio $L_\text{fringe} / L$, and the functional form of $\varepsilon(x)$. Such a study represents a significant numerical effort that we believe is best suited for a future, dedicated paper.}\\  

\begin{table}[H]
    \centering    
    \caption{Investigation of the dependence of the measured mass on fringe boundary condition parameters. The baseline from the paper is $L=128$, $L_\text{fringe}=L$, constant $\varepsilon=10^{-10}$, yielding a mass of $2.06 \pm 0.05$.}
    \label{tab:fringe_tests}
    \begin{tabular}{llc}
        \toprule
        \textbf{Test Type} & \textbf{Parameters} & \textbf{Measured Mass} \\ 
        \midrule
        \textbf{Constant $\varepsilon$} & $\varepsilon=10^{-3}$ & $2.09 \pm 0.04$ \\
        ($L=128, L_\text{fringe}=L$) & $\varepsilon=10^{-5}$ & $2.06 \pm 0.03$ \\
        & $\varepsilon=10^{-10}$ & $2.06 \pm 0.05$ \\
        & $\varepsilon=10^{-15}$ & $1.97 \pm 0.04$ \\
        \midrule
        \textbf{Varying $\varepsilon(x)$ form} & Linear Decay & $2.31 \pm 0.06$ \\
        ($L=128, L_\text{fringe}=L$) & Exponential Decay, $c = 0.05$ & $2.22 \pm 0.07$ \\
        & Exponential Decay, $c = 0.5$ & $2.13 \pm 0.06$ \\
        & Exponential Decay, $c = 1.0$ & $1.99 \pm 0.04$ \\
        \midrule
        \textbf{Varying Lattice Size} & $L=128, L_\text{fringe}=L, \varepsilon=10^{-10}$ & $2.06 \pm 0.05$ \\
        ($L_\text{fringe}=128$) & $L=256, L_\text{fringe}=L/2, \varepsilon=10^{-10}$ & $1.91 \pm 0.02$ \\
        \cline{2-3}
        & $L=128, L_\text{fringe}=L, \varepsilon=10^{-15}$ & $1.97 \pm 0.04$ \\
        & $L=256, L_\text{fringe}=L/2, \varepsilon=10^{-15}$ & $1.91 \pm 0.02$ \\
        \bottomrule
    \end{tabular}
\end{table}

%%%%%%%%%%%%%%%%%%%%%%%%%%%%%%%%

% Bibliography

%% [A] Recommended: using JHEP.bst file
\bibliographystyle{JHEP}
\bibliography{biblio.bib}

\end{document}